\documentclass[11pt]{article}
\usepackage[letterpaper,hmargin=1.3in,vmargin=1.45in]{geometry}
\usepackage{url}
\usepackage{url,color}
\usepackage{graphicx,cite}
\usepackage{epsfig,epstopdf}
\usepackage{enumerate,lineno}
\usepackage{mathrsfs,amsfonts,amsmath,amsthm,amssymb}
\usepackage[colorlinks, citecolor=blue, linkcolor=black]{hyperref}
\usepackage{CJK}
\usepackage{array}

\newcolumntype{C}[1]{>{\centering}p{#1}}
\newcounter{myfootertablecounter}
\newcommand{\FF}{\mathbb{F}}
\newcommand{\F}{\mathbb{F}_2}
\newcommand{\B}{\mathcal {B}}
\newcommand{\N}{\mathcal {N}}
\newcommand{\R}{\mathcal {R}}
\newcommand{\oo}{\overline}
\newtheorem{theorem}{Theorem}

\newtheorem{definition}{Definition}
\newtheorem{example}{Example}
\newtheorem{construction}{Construction}

\newtheorem{remark}{Remark}

\newtheorem{lemma}{Lemma}

\begin{document}

\begin{CJK*}{GBK}{}

\title{Large Sets of Orthogonal Sequences Suitable for Applications in CDMA Systems}
\author{WeiGuo Zhang, ChunLei Xie
\\ ISN Laboratory, Xidian University, Xi'an 710071, China\\
e-mail: zwg@xidian.edu.cn
\and Enes Pasalic \\
University of Primorska, FAMNIT, Koper 6000, Slovenia\\
e-mail: enes.pasalic6@gmail.com }
\date{}

\maketitle

\begin{abstract}
In this paper, we employ the so-called semi-bent functions to achieve significant improvements over currently known methods regarding the number of orthogonal sequences per cell that can be assigned to a regular tessellation of hexagonal cells, typical for certain code-division multiple-access (CDMA) systems. Our initial design method generates a large family of orthogonal sets of sequences derived from vectorial semi-bent functions. A modification of the original approach is proposed to avoid a hard combinatorial problem of allocating several such orthogonal sets to a single cell of a regular hexagonal network, while preserving the orthogonality to adjacent cells. This modification increases the number of users per cell by starting from shorter codewords and then extending the length of these codewords to the desired length. The specification and assignment of these orthogonal sets to a regular tessellation of hexagonal cells have been solved regardless of the parity and size of $m$ (where $2^m$ is the length of the codewords). In particular, when the re-use distance is $D=4$ the number of users per cell is $2^{m-2}$ for almost all $m$, which is twice as many as can be obtained by the best known methods.
\end{abstract}


\section{Introduction}

The design of CDMA systems requires many codewords, both to allow a sufficient number of users in each cell, and to avoid interference arising from the re-use of a codeword in a close cell. A usual way of constructing spreading codes in these systems is to employ correlation-constrained sets of Hadamard matrices \cite{Smith2009,Tang2006,Yang2000} ensuring that the cross-correlation of the rows of different matrices lies in the range $[2^{m/2},2^{\lfloor (m+2)/2 \rfloor}]$. A Hadamard matrix, denoted by $\mathcal {H}$, is an $n \times n$ matrix with elements in $\{+1,-1\}$ satisfying $\mathcal {H}\mathcal {H}^T=nI$, where $I$ is the $n \times n$ identity matrix. Given a set of such matrices $\{\mathcal {H}^{(i)} \}_{i=1}^u$, then denoting by ${\bf r}_j^{(i)}$ the $j$th row of $\mathcal {H}^{(i)}$ the set $\{\mathcal {H}^{(i)} \}_{i=1}^u$ is called correlation-constrained if the inner product $ {\bf r}_{j_1}^{(i_1)}\cdot {\bf r}_{j_2}^{(i_2)}$ ($i_1 \neq i_2$) lies in the range $[-\zeta,\zeta]$. A particular choice of these matrices was considered in \cite{Smith2009,Tang2006,Yang2000} and it was ensured that for $n=2^m$ the value of $\zeta$ lies in the range $[2^{m/2},2^{\lfloor (m+2)/2 \rfloor}]$.

It should be mentioned that apart from synchronous CDMA (S-CDMA) systems treated in this paper, there is an alternative approach known as QS-CDMA which stands for quasi-synchronous CDMA systems \cite{Akansu2007}. For these systems it is of importance that the inner product of two sequences has zero cross-correlation even though one of the sequences is cyclically shifted within a certain range (known as a zero correlation zone). More precisely, for two binary sequences ${\bf x}=x_0,\ldots,x_{n-1}$ and ${\bf y}=y_0,\ldots,y_{n-1}$ of length $n$, the sum $\theta_{{\bf x},{\bf y}}(\tau)=\sum_{j=0}^{n-1}x_j y_{j+\tau}$ should equal to zero for $|\tau| < T$, where $T$ is a positive integer \cite{Sarw1980,Tang2001}.
Loosely synchronized (LS) codes used for QS-CDMA are constructed using Hadamard matrices. These Hadamard matrices can be replaced by the correlation-constrained sets constructed here with the same benefits as for S-CDMA.

Our major goal in this work is to identify construction methods that generate a large set of orthogonal sequences (useful in S-CDMA systems) and to efficiently solve the problem of assigning these sequences to a regular tessellation of hexagonal cells in such a way that the correlation between codewords assigned to adjacent cells is zero, while the correlation between codewords assigned to non-adjacent cells is small. The so-called re-use distance $D$ reflects the ability to use the same codewords in non-adjacent cells that are at distance $D$ from the cell where these codewords have originally been placed. This problem was addressed in \cite{Smith2010}, where the authors employed four cosets of a certain subcode of the first order Reed-Muller code in order to construct 12 sets of orthogonal sequences based on Hadamard matrices and semi-bent functions. In addition, a suitable assignment of these sets into octants or quadrants (see \cite{Smith2010} for definitions) with $D=4$ was specified, and furthermore it was ensured that the inner product of the sequences in the same cell or in adjacent cells is zero. Moreover, it was shown that the inner product of the sequences in non-adjacent cells is at most $2^{\lfloor (m+2)/2 \rfloor}$, and the total number of sequences
assigned was $4 \cdot 2^m$, see \cite{Smith2010}. This approach yielded a significant improvement over other methods due to the specific properties of the Hadamard matrices specified in \cite{Smith2010}.

In this paper, we employ plateaued sequences (whose corresponding Walsh spectra are three-valued) derived from the Maiorana-McFarland class of Boolean functions. These sequences also serve the purpose of constructing large sets of orthogonal sequences which may be efficiently assigned to a regular tessellation of hexagonal cells. We emphasize that we only consider the assignment to a regular hexagonal structure whereas the assignment to irregular networks can be done using heuristic algorithms developed in \cite{Smith2013}. Even though we generally consider plateaued sequences, we essentially use semi-bent sequences whose maximum correlation is $2^{\lfloor (m+2)/2 \rfloor}$ due to the requirement that the correlation of non-orthogonal sequences should be as small as possible.

Our basic design approach, based on the use of vectorial semi-bent functions, generates a large family of sets of mutually orthogonal sequences (within each set) and is also characterized by the property that a great majority of these sets are also mutually orthogonal. To avoid a hard combinatorial problem of assigning several such orthogonal sets to a single cell, while preserving the orthogonality to adjacent cells, this method is combined with the so-called bent concatenation used for extending the length of these codewords to the desired length. This allows us to increase the number of orthogonal sequences per cell without modifying the original assignment found for small suitably chosen $m$. Nevertheless, to achieve an optimal assignment of orthogonal sets when $m$ is even (in terms of the number of users per cell) we employ a slightly modified basic approach which ensures that the number of sequences per cell is $2^{m-2}$ in this case as well. To summarize, compared to the methods in \cite{Smith2010} which for $D = 4$ give $2^{m-3}$ orthogonal sequences per cell for all $m$ (apart from the cases $m = 3, 9$ when their number is $2^{m-2}$ \cite{Smith2010, Smith2012}), our approach can be used for assigning $2^{m-2}$ orthogonal sequences per cell for any $m \neq 4, 5$. The methods described here may also be used for allocating $2^{m-3}$ orthogonal sequences per cell for $D=8$ and almost all $m$.

Finally, the possibility of using our families of orthogonal sets within the framework of orthogonal variable spreading factor (OVSF) codes is also addressed. These codes support variable data rates by shortening the length $2^m$ of the original sequences by some variable factor $2^l$ (where $1 \leq l \leq m-2$), while the orthogonality and low correlation value (when mutually non-orthogonal) of these subsequences is preserved. It turns out that our first modification of the basic method, referred to as Construction~\ref{constr:2} in this paper, ensures that (only) a certain shortening of the codewords satisfies the main characterization of OVSF codes. This eventually leads to an interesting open problem of constructing vectorial bent functions whose restrictions are semi-bent which would essentially give us the possibility to construct OVSF codes whose data rate can be varied within a wider range.

This paper is organized as follows. Some basic notions and concepts related to sequences are introduced in Section~\ref{sec:prel}. In Section~\ref{sec:main}, a design of semi-bent vectorial Boolean functions, upon which a large set of orthogonal semi-bent sequences is derived, is presented. Two modifications of our main method, that allow us to fully specify the assignment of the orthogonal sets of sequences within a regular hexagonal tessellation of cells for (almost) any $m$, are discussed in Section~\ref{sec:modif}. A comparison of our methods to the approach in \cite{Smith2010}, in terms of the number of sequences per cell, is also given here. In Section~\ref{sec:qos}, we discuss the possibility of using our construction techniques for the purpose of generating OVSF codes and quasi-orthogonal sequences. Some concluding remarks are given in Section~\ref{sec:conc}.

\section{Preliminaries}\label{sec:prel}

In this section we present some important notions and tools related to sequences and Boolean functions. Our main tool in the analysis is the Walsh-Hadamard transform.

Let $\mathbb{F}_{2^m}$ and $\F^m$ denote the finite field $GF(2^m)$ and the corresponding vector space, respectively. An $m$-variable Boolean function $f$ is a function from $\F^m$ to $\F$, thus for any fixed $x=(x_1,\ldots,x_m)\in \F^m$ we have $x \overset{f}{\mapsto} f(x) \in \F$. The set of all Boolean functions in $m$-variables is denoted by $\B_m$. For simplicity, we use ``+" and $\sum_i$ to denote the addition operations over $\F^m$ and $\mathbb{F}_{2^m}$.
A Boolean function $f\in \B_m$ is generally represented by its \emph{algebraic normal form}
(ANF):
\begin{eqnarray}
 f(x_1,\ldots,x_m)=\sum_{b\in\F^m}\lambda_b(\prod_{i=1}^m x_i^{b_i}),
\end{eqnarray}
where $\lambda_b\in\F, \ b=(b_1,\ldots,b_m) \in \F^m.$  The algebraic degree of $f(x)$, denoted by $deg(f)$, is the maximal value of $wt(b)$ such that $\lambda_b\neq 0$, where $wt(b)$ denotes the Hamming weight of $b$. $f$ is called an affine function when $deg(f)=1$. An affine function with its constant term equal to zero is called a linear function. For $a=(a_1,\ldots,a_m)\in \F^m$, $b=(b_1,\ldots,b_m)\in \F^m$, the \emph{inner product} of $a$ and $b$ is defined by
\begin{equation}
a\cdot b=\sum_{i=1}^ma_ib_i,
\end{equation}
where addition is performed modulo two. Any linear function on $\F^m$ is defined using the inner product $\omega \cdot x$, where $\omega=(\omega_1,\ldots,\omega_m),\ x=(x_1,\ldots,x_m)\in \F^m$, and each different $\omega$ specifies a distinct linear function. The set of all linear functions in $m$ variables is denoted by $\mathcal{L}_m$, thus $\mathcal{L}_m=\{w \cdot x \mid \omega \in \F^m \}$.

The Walsh-Hadamard transform of $f\in \B_m$ at point $\omega$ is denoted by $W_f(\omega)$ and it is computed as
\begin{eqnarray}
 W_f(\omega)=\sum_{x\in \mathbb {F}_2^m} (-1)^{f(x)+ \omega\cdot x}.
\end{eqnarray}
Let $supp(f)=\{x\in\F^{m}\mid f(x)=1\}$ denote the support of $f$. Then, $f\in \B_m$ is said to be balanced if its output column in the truth table contains equal
number of $0$'s and $1$'s, i.e., $\# supp(f)=2^{m-1}$, or equivalently
\begin{equation}\label{}
W_f(\textbf{0}_m)=0,
\end{equation}
where $\textbf{0}_m$ denotes the all zero vector of length $m$.
Parseval's equation \cite{MacWilliams} states that
\begin{equation}
\sum_{\omega\in \mathbb{F}_2^m}(W_f(\omega))^2=2^{2m}
\end{equation}
and implies that
$
\max\limits_{\omega\in \F^m}|W_f(\omega)|\geq 2^{m/2}.
$
The equality occurs if and only if $f\in \mathcal {B}_m$ is a bent
function \cite{Rothaus}, where $m$ must be even.
A function $f\in \B_m$ satisfying that $W_f(\omega)\in \{\pm 2^{m/2}\}$, for all $\omega \in \mathbb{F}_2^m$, is called bent.

The \emph{sequence} of $f\in \B_m$ is a $(1,-1)$-sequence of length $N=2^m$ defined as
 \begin{equation}
 \oo{f}=\left((-1)^{f(0,\ldots ,0,0)},(-1)^{f(0,\ldots ,0,1)},\ldots, (-1)^{f(1,\ldots ,1,1)}\right).
\end{equation}
The \emph{inner product} of $\oo{f_1}$ and $\oo{f_2}$, denoted by $\oo{f_1}\cdot \oo{f_2}$, is defined by
\begin{equation}
\oo{f_1}\cdot \oo{f_2}=\sum_{i=1}^Nu_iv_i.
\end{equation}
It easily follows that $W_f(\omega)=\oo{f}\cdot\oo{l}$, where $l=\omega\cdot x$.
A $2^m \times 2^m$ Sylvester-Hadamard matrix, denoted
by $\mathcal {H}_m$, is generated by the following recursive relation:

\begin{equation}\label{}
 \mathcal {H}_0=(1)~~~\mathcal {H}_m=\left(
 \begin{array}{cc}
 \mathcal {H}_{m-1} & \mathcal {H}_{m-1} \\
 \mathcal {H}_{m-1} & -\mathcal {H}_{m-1} \\
 \end{array}
 \right),
 ~~m=1,2,\cdots
\end{equation}
Let ${\bf r}_j, 0\leq j \leq 2^m-1$, be the $j$th row of $\mathcal {H}_m$. Then, ${\bf r}_j$ is the
sequence of the linear function $\omega\cdot x$, where $\omega\in \F^m$ corresponds to the binary representation
of the integer $j$. We often call
\begin{equation}\label{}
 \textbf{H} = \{{\bf r}_j\mid 0\leq j \leq 2^m-1\}.
\end{equation}
a set of Hadamard sequences. Clearly,
\begin{equation}\label{}
 \textbf{H} = \{\oo{l}~|~\l\in \mathcal{L}_m\}.
\end{equation}

For $f_1$, $f_2\in \B_m$, let $\oo{f_1}=(u_1,\ldots,u_N)$ and $\oo{f_2}=(v_1,\ldots,v_N)$
be the sequences of $f_1$ and $f_2$, respectively. The \emph{componentwise product} of $\oo{f_1}$ and $\oo{f_2}$
is defined by
 \begin{equation}
\oo{f_1}*\oo{f_2}=(u_1v_1,\ldots,u_Nv_N).
\end{equation}
Obviously,
 \begin{equation}
 \oo{f_1}* \oo{f_2}= \oo{f_1+f_2}.
\end{equation}

\begin{definition}
Let $f_1,f_2\in \B_m$. $\oo{f_1}$ and $\oo{f_2}$ are orthogonal, denoted by $\oo{f_1}\bot\oo{f_2}$, if
\begin{equation}
\oo{f_1}\cdot\oo{f_2}=\sum_{x \in \F^m}(-1)^{f_1(x)+ f_2(x)}=0.
\end{equation}
We call
\begin{equation}\label{}
 S=\{ \oo{{f_i}}~|~f_i\in \B_m, i=1,2,\ldots, \kappa\}
\end{equation}
 a set of orthogonal sequences of cardinality $\kappa$ if the sequences in $S$ are pairwise orthogonal.
 Let $S_1$ and $S_2$ be two sets of orthogonal sequences. $S_1$ and $S_2$ are orthogonal to each other, denoted by $S_1 \bot S_2$,
 if $\oo{f_1}\cdot\oo{f_2}=0$ always holds for any $\oo{{f_1}}\in S_1$ and $\oo{f_2}\in S_2$.
\end{definition}
\noindent Noticing that
\begin{eqnarray}
W_{f_1+f_2}(\textbf{0}_m)=0 &\Leftrightarrow& \sum_{x\in \F^m}(-1)^{f_1(x)+f_2(x)}=0 \nonumber\\
&\Leftrightarrow& \sum_{x\in \F^m}(-1)^{f_1(x)}(-1)^{f_2(x)}=0 \nonumber \\
&\Leftrightarrow& \oo{f_1}\cdot\oo{f_2}=0,
\end{eqnarray}
 we obtain the following simple but important characterization of orthogonal sequences.
\begin{lemma}\label{orth}
Let $f_1,f_2\in \B_m$. Then $\oo{f_1} \bot \oo{f_2}$ if and only if $W_{f_1+f_2}(\textnormal{\textbf{0}}_m)=0$.
\end{lemma}
\noindent For any two different linear functions $l,l' \in \mathcal{L}_m$,
$W_{l+l'}(\textnormal{\textbf{0}}_m)=0$. Then $\oo{l}\bot \oo{l'}$ always holds, which implies
$\textbf{H}$ is a set of orthogonal sequences.

The notion of plateaued functions, also called three-valued spectra functions, was introduced by Zheng and Zhang \cite{Zheng2001} to facilitate the design of cryptographically good functions.
These functions include semi-bent functions \cite{Chee1994} as a proper subset.

\begin{definition}\label{def:1}
A function $f\in \B_m$ is called a plateaued function if $W_f(\alpha)\in \{0, \pm 2^\lambda\}$ for any $\alpha \in \F^m$, where $\lambda\geq m/2$ is a positive integer.
When $\lambda=\lfloor (m+2)/2 \rfloor$, $f$ is called a semi-bent function. $\oo{f}$ is called a plateaued sequence (respectively semi-bent sequence)
if $f$ is a plateaued function (respectively semi-bent function).
\end{definition}
\noindent Plateaued functions and bent functions can be obtained by the \emph{Maiorana-McFarland} construction method. The Maiorana-McFarland class of functions is defined as follows.

\begin{definition}\label{def:MM} For any positive
integers $s$, $t$ such that $m=s+t$, an Maiorana-McFarland function $f\in \B_m$ is defined by
\begin{equation} \label{eq:MM}
f(y, x)=\phi(y)\cdot x\oplus \tau(y), ~~~ y\in\F^s, \; x\in \F^t,
\end{equation}
 where $\phi$ is any mapping
from $\F^s$ to $\F^t$ and $\tau\in \B_s$.
\end{definition}
\noindent When $s\leq t$ and $\phi$ is injective, the Maiorana-McFarland functions are plateaued.
In particular, when $s=t$ and $\phi$ is bijective, we get the Maiorana-McFarland class of bent functions.

\begin{definition}\label{def:3}
An $m$-variable $t$-dimensional vectorial function is a mapping $F: \F^m\mapsto
\F^t$, which can also be viewed as a collection of $t$ Boolean
functions so that $F(x)=(f_1,\ldots,f_t)$, where $f_1,\ldots,f_t \in \B_m$.
$F$ is called a vectorial plateaued function if any nonzero linear combination of the component functions $f_1,\ldots,f_t$ is a plateaued Boolean function with three-valued Walsh spectra $\{0,\pm 2^{\lambda}\}$.
When $\lambda=\lfloor (m+2)/2 \rfloor $, $F$ is called a vectorial semi-bent function.
$F$ is called a vectorial bent function if any nonzero linear combination of $f_1,\ldots,f_t$ is a bent function with two-valued Walsh spectra $\{\pm 2^{m/2}\}$.
\end{definition}

\section{Vectorial semi-bent functions and sets of orthogonal sequences}\label{sec:main}

In this section we derive a family of sets of orthogonal sequences from a single vectorial plateaued (semi-bent) function $F: \F^m\mapsto \F^t$. The main idea of finding large sets of mutually zero-correlated sequences can be described as follows.
Once the function $F$, with the design parameters $s$ and $t$ so that $m=s+t$, has been specified, then $2^{2t}$ sets of orthogonal sequences are constructed by using linear combinations of its Boolean components along with the addition of linear functions.
Each of these $2^{2t}$ sets will contain $2^s$ mutually orthogonal sequences and furthermore each such a set is orthogonal to a great majority of the remaining sets. However, a careful placement in the cell cluster is required to achieve orthogonality of adjacent cells or possibly the orthogonality within the same cell if several such sets are assigned to a single cell since not all of these sets are mutually orthogonal to each other.

\subsection{Construction of a large family of sets of orthogonal sequences}

The construction below efficiently employs the Marioana-McFarland class of functions, taking full advantage of its affine equivalence class for our purpose. In the sequel, we frequently identify the vector space $\F^t$ with the finite field $\FF_{2^t}$ through the canonical linear isomorphism $\pi: \FF_{2^t}\mapsto \F^t$ so that $\pi(b_1+b_2\gamma+\cdots+b_{t}\gamma^{t-1})=(b_1,b_2,\ldots,b_{t})$, where $\gamma$ is a primitive element in $\FF_{2^t}$ and therefore $\{1,\gamma, \ldots, \gamma^{t-1}\}$ is a polynomial basis of $\FF_{2^t}$ over $\F$ implying that any $\alpha \in \FF_{2^t}$ can be expressed as $\alpha=b_1+b_2\gamma+\cdots+b_{t}\gamma^{t-1}$ for suitably chosen $b_i \in \F$.
\begin{construction}\label{constr:1}
Let $m$, $s$, and $t$ be three positive integers satisfying $m=s+t$, $s<t$.
For $y\in \F^s$ let $[y]$ denote the integer representation of $y$, i.e., $[y]=\sum_{j=1}^{s}y_j2^{j-1}$ for a fixed $y\in \F^s$. For $i=1, \ldots, t$ define a collection of Marioana-McFarland functions
 \begin{equation}\label{eq:g_i}
f_i(y,x)=\phi_i(y)\cdot x, ~~ x\in \F^t, \; y\in \F^s,
 \end{equation}
where $\phi_i:\F^s \rightarrow \F^t$ is an injective mapping defined by $\phi_i(y)=\pi(\gamma^{[y]+i})$.
We define a vectorial function $F: \F^{m} \mapsto \F^{t}$ by
 \begin{equation}
F(y,x)=(f_1(y,x),\cdots, f_t(y,x)).
 \end{equation}
For any fixed $\alpha\in \F^t$, let
 \begin{equation}\label{eq:Lalpha}
L_{\alpha}(y,x)=\{l_\beta(y,x)=(\beta,\alpha)\cdot (y,x)~|~\beta\in \F^s\}.
 \end{equation}
For any $c\in {\F^t}$, let $f_{c}(y,x)=c\cdot F(y,x)=c_1f_1(y,x)+ \cdots + c_t f_t(y,x)$. We construct $2^{2t}$ disjoint sets of sequences as follows:
\begin{equation}\label{eq:Scl}
S_{c,\alpha}=\{\oo{f_c} * \oo{l}~|~l\in L_\alpha\},~~\mathrm{for}~c, \alpha\in {\F^t}.
\end{equation}
\end{construction}

\begin{remark}\label{remark:1}
We list several important observations related to Construction~\ref{constr:1}.
\begin{enumerate}[i)]

 \item For $c,c'\in \F^{t}$, $f_c+f_{c'}=f_{c+c'}$.

 \item $S_{c,\alpha}=\{\oo{f_{c} + l}~|~l\in L_{\alpha}\}$; $\# S_{c,\alpha}=2^s$.

 \item For any $\alpha \in \F^t$, let $H_{\alpha}=S_{\textbf{0}_t,\alpha}=\{\oo{l}~|~l\in L_\alpha\}=\{\oo{{{(\beta \cdot y + \alpha \cdot x )}}}~|~\beta \in \F^s \}$.
 Then $\textbf{H}=\cup_{\alpha\in \F^t}H_{\alpha}$.
\end{enumerate}
\end{remark}

The above construction gives a general design of three-valued plateaued sequences whose spectra are $\{0,\pm 2^t \}$. In order to minimize the correlation between non-orthogonal sequences it is desirable to choose $t$ as small as possible. However, due to the design $m=s+t$ and $s <t$, the minimum value of $t$ is $\lfloor (m+2)/2 \rfloor$ which essentially corresponds to semi-bent sequences.

\begin{lemma}\label{lemma:2}
The function $F: \F^{m} \mapsto \F^{t}$ defined in Constrution \ref{constr:1} is a vectorial plateaued function, more precisely any nonzero linear combination of its component functions is a plateaued Boolean function with spectra $\{0,\pm 2^t\}$.
Furthermore, when $s=\lfloor (m-1)/2\rfloor$ and $t=\lfloor (m+2)/2 \rfloor$, i.e.,
\begin{eqnarray}
t=\left\{ \begin{array}{ll}
 (m+1)/2, & \textrm{$m$ is odd}\\
 (m+2)/2, & \textrm{$m$ is even},
\end{array} \right.
\end{eqnarray}
$F$ is a vectorial semi-bent function.
\end{lemma}

\begin{proof}
For a nonzero $c = (c_1,\ldots, c_t)\in \F^t$ and for $y \in \F^s$, $x \in \F^t$ we can write
\begin{eqnarray}\label{eq:fc}
f_c(y,x)&=&\sum_{i=1}^t c_if_i(y,x)=\sum_{i=1}^t c_i\phi_i(y)\cdot x \nonumber\\
&=&\Big (\sum_{i=1}^t c_i\pi(\gamma^{[y]+i})\Big )\cdot x \nonumber\\
&=& \pi\Big(\sum_{i=1}^t c_i\gamma^{[y]+i}\Big )\cdot x
\end{eqnarray}
where the last equality is due to the fact that $\pi$ is a linear isomorphism. Noticing that
 $\gamma$ is primitive in $\FF_{2^t}$, there exists $0\leq i_c \leq 2^t-2$ such that
$\gamma^{i_c}=c\cdot(1, \ldots, \gamma^{t-1})$. Thus,
 \begin{equation}
f_c(y,x) = \pi(\gamma^{i_c+[y]})\cdot x.
\end{equation}
For any $(\beta,\alpha) \in \F^s \times \F^t$, we have
\begin{align}\label{eq:orthog}
W_{ f_{c}}(\beta,\alpha) &=
\sum_{(y,x)\in \F^m}(-1)^{f_{c}(y,x) + \beta \cdot y + \alpha \cdot x}\nonumber\\
&= \sum_{y\in \F^s}(-1)^{\beta \cdot y } \sum_{x\in \F^t}(-1)^{ \pi(\gamma^{[y]+i_{c}})\cdot x +\alpha \cdot x}\nonumber\\
&= \left\{ \begin{array}{ll}
0, & \textrm{if $\pi^{-1}(\alpha)\notin \{\gamma^{[y]+i_{c+e}}~|~y\in \F^s\}$}\\
\pm 2^t, & \textrm{otherwise}.
\end{array} \right.
 \end{align}
 The last equality comes from the fact that $\pi$ is injective and thus there might exist a unique $y \in \F^s$ such that $\pi(\gamma^{[y]+i_{c}})=\alpha$ in which case $\sum_{x\in \F^t}(-1)^{ \pi(\gamma^{[y]+i_{c}})\cdot x +\alpha \cdot x}=2^t$, otherwise this sum equals zero for any $y \in \F^s$.
 By Definition \ref{def:3}, when $t=\lfloor (m+2)/2 \rfloor $, $F$ is a vectorial semi-bent function.
\end{proof}

The important special case of orthogonal sets of \emph{semi-bent} sequences is given below.

\begin{theorem}\label{th1}
Let $m=s+t$ with $s=\lfloor (m-1)/2\rfloor$ and $t=\lfloor (m+2)/2 \rfloor$.
For $c, \alpha\in {\F^t}$, let the sets of sequences $S_{c,\alpha}$ be defined by (\ref{eq:Scl}) as in Construction \ref{constr:1}.
Then we have
\begin{enumerate}[i)]
\item All the sequences in $S_{c,\alpha}$ are semi-bent sequences;

\item For any fixed $c, \alpha\in {\F^t}$, $S_{c,\alpha}$ is a set of orthogonal sequences;

\item
 For any $c\neq e$, $S_{c,\alpha}\bot S_{e,\delta}$ if and only if
 \begin{equation}\label{condition3}
 \pi^{-1}(\alpha+\delta)\notin \{\gamma^{[y]+i_{c+e}}~|~y\in \F^s\},
 \end{equation}
 where $\gamma^{i_c}=c\cdot(1, \ldots, \gamma^{t-1})$. In particular, $S_{c,\alpha}\bot S_{e,\delta}$ when $c\neq e$ and $\alpha=\delta$.
 In addition, for a fixed $c \in \F^t$, $S_{c,\alpha}\bot S_{c,\alpha'}$ whenever $\alpha \neq \alpha'$.

 \item For any fixed $e, \delta\in {\F^t}$, there exist $M$ sets in
 $\{S_{c,\alpha}~|~c, \alpha\in {\F^t}\}$ which are orthogonal to $S_{e,\delta}$, where
\begin{eqnarray}\label{M}
M=\left\{ \begin{array}{ll}
 2^m+2^{(m-1)/2}-1, & \textrm{$m$ is odd}\\
 3\cdot 2^m+2^{m/2-1}-1, & \textrm{$m$ is even}.
\end{array} \right.
\end{eqnarray}
\end{enumerate}
\end{theorem}

\begin{proof}
$i)$
By Lemma \ref{lemma:2}, for any $c,\alpha\in \F^t$, all the sequences in $S_{c,\alpha}$ are semi-bent sequences.

$ii)$
For any $\beta,\beta'\in \F^s$ with $\beta \neq \beta'$, let $\oo{f_c} * \oo{l_\beta}$, $\oo{f_c} * \oo{l_{\beta'}}\in S_{c,\alpha}$.
Note that $l_\beta + l_{\beta'}$ is a nonzero linear function. Then,
\begin{eqnarray}
(\oo{f_c} * \oo{l_\beta})\cdot (\oo{f_c} * \oo{l_{\beta'}})
&=&\oo{{(f_c + l_\beta)}}\cdot \oo{{(f_c + l_{\beta'})}}\nonumber\\
&=&\sum_{(y,x)\in \F^m} (-1)^{{(l_\beta + l_{\beta'})}(y,x)}\nonumber\\
&=&0.
\end{eqnarray}
This proves that $S_{c,\alpha}$ is a set of orthogonal sequences.

$iii)$
For $c$, $\alpha \in {\F^t}$, we define
 \begin{equation}
\Gamma_{c,\alpha}=f_{c} + L_{\alpha}=\{f_{c} + l~|~l\in L_{\alpha}\},
 \end{equation}
where $L_{\alpha}$ is given by (\ref{eq:Lalpha}). Then, for $e, \delta\in {\F^t}$, let $\Gamma_{e,\delta}=f_{e} + L_{\delta}=\{f_{e} + l~|~l\in L_{\delta}\}$, where $L_{\delta}=\{l_\theta=(\theta,\delta)\cdot (y,x)~|~\theta\in \F^s\}$.
 For any $f_{c,\alpha}\in \Gamma_{c,\alpha}$, $f_{e,\delta}\in \Gamma_{e,\delta}$, we consider the following
 two cases.

 \noindent {\bf Case 1:} $c\neq e$. Then,
 \begin{eqnarray}
 & & f_{c,\alpha}(y,x) + f_{e,\delta}(y,x) \nonumber\\
 &=& (f_c(y,x)+ f_e(y,x)) + (l_\beta(y,x)+l_\theta(y,x)) \nonumber\\
 &=& f_{c+e}(y,x) + (\beta+\theta)\cdot y + (\alpha+\delta)\cdot x \nonumber\\
 &=& \phi_{i_{c+e}}(y)\cdot x + (\beta+\theta)\cdot y + (\alpha+\delta)\cdot x \nonumber\\
 &=& \left(\pi(\gamma^{[y]+i_{c+e}})+\alpha+\delta\right)\cdot x + (\beta+\theta)\cdot y.
 \end{eqnarray}
Therefore,
 \begin{eqnarray}
& & W_{ f_{c,\alpha} + f_{e,\delta}}(\textbf{0}_m)\nonumber\\
&=& W_{ f_{c+e}}(\beta+\theta, \alpha+\delta)\nonumber \\
&=& \sum_{(y,x)\in \F^m}(-1)^{f_{c+e}(y,x) + (\beta+\theta)\cdot y + (\alpha+\delta)\cdot x}\nonumber\\
&=& \sum_{y\in \F^s}(-1)^{(\beta+\theta)\cdot y } \sum_{x\in \F^t}(-1)^{ \left(\pi(\gamma^{[y]+i_{c+e}})+\alpha+\delta\right)\cdot x}\nonumber\\
&=& \left\{ \begin{array}{ll}
0, & \textrm{if {\small$\pi^{-1}(\alpha+\delta)\notin \{\gamma^{[y]+i_{c+e}}~|~y\in \F^s\}$}}\\
\pm 2^t, & \textrm{otherwise}.
\end{array} \right.
 \end{eqnarray}
By Lemma \ref{orth} and (\ref{eq:orthog}), $S_{c,\alpha}\bot S_{e,\delta}$ if and only if
 \begin{equation}
 \pi^{-1}(\alpha+\delta)\notin \{\gamma^{[y]+i_{c+e}}~|~y\in \F^s\}.
 \end{equation}
Note that $\pi$ is an injective mapping. Then we have
 $\#\{\alpha\in \F^t~|~\pi^{-1}(\alpha+\delta)=\emptyset \}= 2^t-2^s.$
 Namely, there exist $2^t-2^s$ vectors $\alpha\in\F^t$ such that
$W_{ f_{c,\alpha} + f_{e,\delta}}(\textbf{0}_m) = 0$.\newline \newline
\noindent {\bf Case 2:} $c=e$. Then,
 \begin{eqnarray}
 f_{c,\alpha} + f_{e,\delta} = l_\beta+l_\theta.
 \end{eqnarray}
 Note that $l_\beta+l_\theta\neq 0$ always holds when $\alpha\neq \delta$.
Therefore, there exist
 $(2^t-1)$ vectors $\delta\in\F^t$ such that
$W_{ f_{c,\alpha} + f_{e,\delta}}(\textbf{0}_m) = 0$.

$iv)$
Combining the two cases above, we have
\begin{eqnarray}\label{eq:nmborth}
 & &\sharp\{\Gamma_{c,\alpha}~|~W_{ f_{c,\alpha} + f_{e,\delta}}(\textbf{0}_m)= 0,~c\in \F^t,~\alpha\in \F^t\}\nonumber\\
 &=&(2^t-1)(2^t-2^s)+(2^t-1),
\end{eqnarray}
where the first term accounts for the case $c \neq e$ (for a fixed $c$ there are $2^t-1$ choices for $e$) and the second term regards the case $c=e$.
Note that $S_{c,\alpha}=\{\oo{f}~|~f\in \Gamma_{c,\alpha}\}$.
By (\ref{eq:nmborth}), there exist $(2^{2t}-2^{m}+2^{s}-1)$ sets in
 $\{S_{c,\alpha}~|~c, \alpha\in {\F^t}\}$ that are orthogonal to $S_{e,\delta}$.
 Since $s=\lfloor (m-1)/2\rfloor$ and $t=\lfloor (m+2)/2 \rfloor$, we get
 \begin{eqnarray}
(s,t)=\left\{ \begin{array}{ll}
 (\frac{m-1}{2},\frac{m+1}{2}), & \textrm{$m$ is odd}\\
 (\frac{m-2}{2},\frac{m+2}{2}), & \textrm{$m$ is even}
\end{array} \right.
\end{eqnarray}
which implies that (\ref{M}) holds replacing $s$ and $t$ in $M=2^{2t}-2^{m}+2^{s}-1$.
\end{proof}

\begin{remark}\label{remark:2}

There are certain observations which need to be emphasized.
\begin{enumerate}[i)]

 \item Let $(c,\alpha)\neq (c',\alpha')$. $S_{c,\alpha}\bot S_{c',\alpha'}$ if and only if there exist $\xi\in S_{c,\alpha}$ and $\xi'\in S_{c',\alpha'}$ such that $\xi \bot \xi'$.

 \item More generally, the function $F$ defined in Construction \ref{constr:1} is a vectorial plateaued function. For any fixed $e, \delta\in {\F^t}$, there exist $(2^{2t}-2^{m}+2^{s}-1)$ many sets in
 $\Omega=\{S_{c,\alpha}~|~c, \alpha\in {\F^t}\}$ that are orthogonal to $S_{e,\delta}$;
\end{enumerate}
\end{remark}

\subsection{Illustrating the partition of orthogonal sets and their cell assignment}

In \cite{Smith2010}, the authors considered the assignment of the rows of orthogonal Hadamard matrices to a lattice of regular hexagonal cells. In what follows, we use Construction~\ref{constr:1} to illustrate a similar assignment for the case $m=5$ (with $s=2$, $t=3$), when each sequence is a codeword of length 32. The assignment to regular hexagonal cells is depicted in Figure~\ref{fig:m=5}, and the re-use distance equals $D=8$ in this case, see for instance the placement of the sets $S_{000,000}$. Furthermore, any cell contains $2^{s}=2^{m-3}=4$ orthogonal sequences. The whole procedure of assigning these sets to a lattice of regular hexagonal cells is discussed in the example below.
\begin{example}\label{ex:m=5}
Let $m=5$, $s=2$, and $t=3$. Let $\gamma\in \FF_{2^3}$ be a root of the primitive polynomial $z^3+z+1$.
We have $\pi(1)=100$, $\pi(\gamma)=010$, $\pi(\gamma^2)=001$, $\pi(\gamma^3)=110$,
$\pi(\gamma^4)=011$, $\pi(\gamma^5)=111$, $\pi(\gamma^6)=101$, $\pi(0)=000$.
For $y\in \F^2$, $x\in \F^3$, a vectorial semi-bent function $F: \F^{5} \mapsto \F^{3}$ is constructed as
 \begin{equation*}
F(y,x)=(f_1, f_2, f_3),
 \end{equation*}
where
\begin{align}
f_1(y,x) &= \pi(\gamma^{[y]+1})\cdot x \nonumber\\
&=\oo{y}_1\oo{y}_2x_2 +\oo{y}_1y_2x_3 \nonumber\\
&+{y}_1\oo{y}_2(x_1+x_2) + y_1y_2(x_2+x_3),
\\
f_2(y,x)& = \pi(\gamma^{[y]+2})\cdot x \nonumber\\
&=\oo{y}_1\oo{y}_2x_3 +\oo{y}_1y_2(x_1+x_2) \nonumber\\
&+{y}_1\oo{y}_2(x_2+x_3) + y_1y_2(x_1+x_2+x_3),
\\
f_3(y,x)&= \pi(\gamma^{[y]+3})\cdot x \nonumber\\
&=\oo{y}_1\oo{y}_2(x_1+x_2)+\oo{y}_1y_2(x_2+x_3) \nonumber\\
&+{y}_1\oo{y}_2(x_1+x_2+x_3) + y_1y_2(x_1+x_3),
\end{align}
using the notation $\oo{y}_i=1+y_i$.
Now, let $f_c(y,x)=c\cdot F(y,x)$, for $c\in \F^3$, so that
\begin{center}
\footnotesize{
\begin{tabular}{|c|c|c|c|c|c|c|c|c|c|c|c|c|c|}
 \hline $c$ &000&001&010&011&100&101&110&111
 \\\hline $i_c$ &- &3 &2 &5 &1 &0 &4 &6
\\ \hline
\end{tabular}
}
\end{center}
Specifying the sequences by only using the signs instead of +1, -1, we have
{
\begin{eqnarray*}
\oo{{f_{000}}}&=&\emph{\texttt{(++++++++++++++++++++++++++++++++)}}\\
\oo{{f_{100}}}&=&\emph{\texttt{(++--++--+-+-+-+-++----+++--++--+)}}\\
\oo{{f_{010}}}&=&\emph{\texttt{(+-+-+-+-++----+++--++--++--+-++-)}}\\
\oo{{f_{001}}}&=&\emph{\texttt{(++----+++--++--++--+-++-+-+--+-+)}}\\
\oo{{f_{110}}}&=&\emph{\texttt{(+--++--++--+-++-+-+--+-+++++----)}}\\
\oo{{f_{011}}}&=&\emph{\texttt{(+--+-++-+-+--+-+++++----++--++--)}}\\
\oo{{f_{111}}}&=&\emph{\texttt{(+-+--+-+++++----++--++--+-+-+-+-)}}\\
\oo{{f_{101}}}&=&\emph{\texttt{(++++----++--++--+-+-+-+-++----++)}}
\end{eqnarray*}
}

For any fixed $\alpha \in \F^3$, let $L_\alpha=\{\beta \cdot y + \alpha \cdot x \mid \beta\in \F^2\}$. Let $\Omega=\{S_{c,\alpha}~|~c,\alpha\in \F^t\}=\{\oo{f_{c} + l}~|~l\in L_{\alpha}\}$. Notice that $\#\Omega =2^{2t}=64$ and each $S_{c,\alpha}$ is of cardinality $2^s=4$.
Let $H_{\alpha}=S_{000,\alpha}=\{\oo{l}~|~l\in L_\alpha\}=\{\oo{{{(\beta \cdot y + \alpha \cdot x )}}}~|~\beta \in \F^2 \}$.
 Then the sequences that belong to $H_{\alpha}$, for different $\alpha\in \F^3$, actually correspond to a partition of the Hadamard matrix $\mathcal {H}_5$ into 8 parts (see Table \ref{partition}), i.e.,
\begin{equation}\label{}
 \textbf{H}=\bigcup_{\alpha\in \F^3}H_\alpha.
\end{equation}
\begin{table}[!tbp]
\caption{A partition of the Hadamard matrix $\mathcal {H}_5$ into 8 parts}
\label{partition}
\begin{center}
\small{
\begin{tabular}{|c|c|c|}
 \hline
$H_{000}$
&\texttt{++++++++++++++++++++++++++++++++}\\
&\texttt{++++++++--------++++++++--------}\\
&\texttt{++++++++++++++++----------------}\\
&\texttt{++++++++----------------++++++++}\\\hline
$H_{100}$
&\texttt{++++----++++----++++----++++----}\\
&\texttt{++++--------++++++++--------++++}\\
&\texttt{++++----++++--------++++----++++}\\
&\texttt{++++--------++++----++++++++----}\\\hline
$H_{010}$
&\texttt{++--++--++--++--++--++--++--++--}\\
&\texttt{++--++----++--++++--++----++--++}\\
&\texttt{++--++--++--++----++--++--++--++}\\
&\texttt{++--++----++--++--++--++++--++--}\\\hline
$H_{001}$
&\texttt{+-+-+-+-+-+-+-+-+-+-+-+-+-+-+-+-}\\
&\texttt{+-+-+-+--+-+-+-++-+-+-+--+-+-+-+}\\
&\texttt{+-+-+-+-+-+-+-+--+-+-+-+-+-+-+-+}\\
&\texttt{+-+-+-+--+-+-+-+-+-+-+-++-+-+-+-}\\\hline
$H_{110}$
&\texttt{++----++++----++++----++++----++}\\
&\texttt{++----++--++++--++----++--++++--}\\
&\texttt{++----++++----++--++++----++++--}\\
&\texttt{++----++--++++----++++--++----++}\\\hline
$H_{011}$
&\texttt{+--++--++--++--++--++--++--++--+}\\
&\texttt{+--++--+-++--++-+--++--+-++--++-}\\
&\texttt{+--++--++--++--+-++--++--++--++-}\\
&\texttt{+--++--+-++--++--++--++-+--++--+}\\\hline
$H_{111}$
&\texttt{+--+-++-+--+-++-+--+-++-+--+-++-}\\
&\texttt{+--+-++--++-+--++--+-++--++-+--+}\\
&\texttt{+--+-++-+--+-++--++-+--+-++-+--+}\\
&\texttt{+--+-++--++-+--+-++-+--++--+-++-}\\\hline
$H_{101}$
&\texttt{+-+--+-++-+--+-++-+--+-++-+--+-+}\\
&\texttt{+-+--+-+-+-++-+-+-+--+-+-+-++-+-}\\
&\texttt{+-+--+-++-+--+-+-+-++-+--+-++-+-}\\
&\texttt{+-+--+-+-+-++-+--+-++-+-+-+--+-+}\\ \hline
\end{tabular}
}
\end{center}
\end{table}
Obviously, the above sets correspond to $f_c=0$, that is $c=0$.
If $\oo{f_{c}}\bot \oo{l}$ for any $\oo{l}\in H_{\alpha}$, then we say that $\oo{f_{c}}$ and $H_{\alpha}$ are orthogonal, denoted by
 $\oo{f_{c}}\bot H_{\alpha}$. In Table~\ref{tab:m=5}, we depict the orthogonality between $\oo{f_{c}}$ and $H_{\alpha}$, where $c,\alpha \in \F^3$.
The reason why we do not sort the sequences in Table~\ref{tab:m=5} in lexicographic order is that the process of determining the orthogonal sets
becomes easier and Table~\ref{tab:m=5} has cyclic structure (the sorting depends on the finite field representation through $\pi$).
\begin{table}[t]
{
\caption{Orthogonality between $\oo{{f_{c}}}$ and $H_{\alpha}$}
\label{tab:m=5}
\begin{center}
{\footnotesize
\begin{tabular}{|c|C{0.5cm}|C{0.5cm}|C{0.5cm}|C{0.5cm}|C{0.5cm}|C{0.5cm}|C{0.5cm}|c| }
\hline
 &$H_{000}$~~&$H_{100}$&$H_{010}$&$H_{001}$&$H_{110}$&$H_{011}$&$H_{111}$&$H_{101}$\\ \hline
$\oo{{f}_{000}}$& & $\bot$ & $\bot$ & $\bot$ & $\bot$ & $\bot$ & $\bot$ & $\bot$ \\\hline
$\oo{{f}_{100}}$& $\bot$ & $\bot$ & & & & & $\bot$ & $\bot$ \\\hline
$\oo{{f}_{010}}$& $\bot$ & $\bot$ & $\bot$ & & & & & $\bot$ \\\hline
$\oo{{f}_{001}}$& $\bot$ & $\bot$ & $\bot$ & $\bot$ & & & & \\\hline
$\oo{{f}_{110}}$& $\bot$ & & $\bot$ & $\bot$ & $\bot$ & & & \\\hline
$\oo{{f}_{011}}$& $\bot$ & & & $\bot$ & $\bot$ & $\bot$ & & \\\hline
$\oo{{f}_{111}}$& $\bot$ & & & & $\bot$ & $\bot$ & $\bot$ & \\\hline
$\oo{{f}_{101}}$& $\bot$ & & & & & $\bot$ & $\bot$ & $\bot$ \\\hline
\end{tabular}
}
\end{center}}
\end{table}

Next we briefly discuss the orthogonality between
 $S_{c,\alpha}=\{\oo{f_c} * \oo{l}~|~\oo{l}\in H_\alpha\}$ and $S_{c',\alpha'}=\{\oo{f_{c'}} * \oo{l'}~|~\oo{l'}\in H_{\alpha'}\}$ in general.
Notice that
\begin{eqnarray}
(\oo{f_c} * \oo{l})\cdot (\oo{f_{c'}} * \oo{l'})&=&\oo{f_c+l} \cdot \oo{f_{c'}+l'}\nonumber\\
&=&\oo{f_c+f_{c'}} \cdot \oo{l+l'}\nonumber\\
&=&\oo{f_{c+c'}} \cdot \oo{l+l'}\nonumber\\
&=&\oo{f_{c+c'}} \cdot (\oo{l}* \oo{l'}),
\end{eqnarray}
and $H_{\alpha}*H_{\alpha'}=H_{\alpha+\alpha'}$, where
\begin{equation}\label{}
 H_{\alpha}*H_{\alpha'}=\{\oo{l}*\oo{l'}~|~\oo{l}\in H_{\alpha}, \oo{l'}\in H_{\alpha'}\}.
\end{equation}
Furthermore, since $\oo{l}* \oo{l'}\in H_{\alpha}*H_{\alpha'}$,
then
\begin{equation}\label{eq:orthog_check}
S_{c,\alpha}\bot S_{c',\alpha'} \Leftrightarrow \oo{{f}_{c+c'}}\bot H_{\alpha+\alpha'}.
\end{equation}
In this example, by Theorem~\ref{th1}, $S_{c,\alpha}\bot S_{c',\alpha'}$ if and only if one of the following conditions holds:
\begin{itemize}
 \item $c=c'$ and $\alpha\neq \alpha'$
 \item $c\neq c'$ and $\alpha=\alpha'$
 \item $c\neq c'$ and $\alpha\neq \alpha'$, there exists $i\in\{0,1,2\}$ such that $\pi^{-1}(c+c')=\pi^{-1}(\alpha+\alpha')\cdot \gamma^i$
\end{itemize}
For instance, consider $S_{c',\alpha'}=S_{011,101}$ (colored green) and its orthogonal sets denoted by $S_{011,101}^{\perp}$.
Let $\R$ contain all the sets $S_{c,\alpha}$ (where $c\neq 011$ and $\alpha \neq 101$) such that $$c+011 \in \{v_1,v_2,\alpha+101\},$$
where $v_1, v_2$
and $\alpha+101$ are three clockwise consecutive vectors in the circle in Figure \ref{fig:1}.
\begin{figure}[t]
\centering
\includegraphics[scale=0.6]{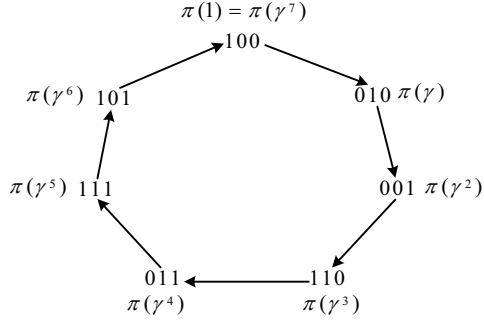}
\caption{The one-to-one relationship between $\F^m$ and $\mathbb{F}_{2^m}$}
\label{fig:1}
\end{figure}
Then,
\begin{align}
S_{011,101}^{\perp}&=\{S_{011, \alpha}~|~\alpha \in \F^3, \alpha \neq 101\} \nonumber\\
&~~~~~~~\cup \{S_{c, 101}~|~c\in \F^3, c \neq 011\} \cup \R.
\end{align}
There are $2^m+2^{(m-1)/2}-1=35$ orthogonal sets to $S_{011,101}$ and $14=7+7$ of these come from the first two sets (the case $c \neq c'$ or $\alpha \neq \alpha'$). This implies that $\# \R =21$.
For instance, starting with $(v_1,v_2,\alpha+101)=(100,010,001)$, which implies $\alpha=100$, we uniquely identify 3 vectors $c$ so that $c + 011 \in \{100,010,001\}$, namely $c \in \{111,001,010\}$. Therefore, $S_{111,100},S_{001,100},S_{010,100} \in \R$. Continuing this way, by taking a new triple of clockwise consecutive vectors on the circle $(v_1,v_2,\alpha+101)=(010,001, 110)$ we would get three more sets $S_{c,010}$ that belong to $\R $. Then, completing the full cycle $21=7 \times 3$ pairs $(c,\alpha)$ are identified.

The assignment of the sets $S_{c,\alpha}$ for the case $m=5$ is depicted in Figure~\ref{fig:m=5}.
\begin{figure}[t]
\centering
\includegraphics[scale=0.6]{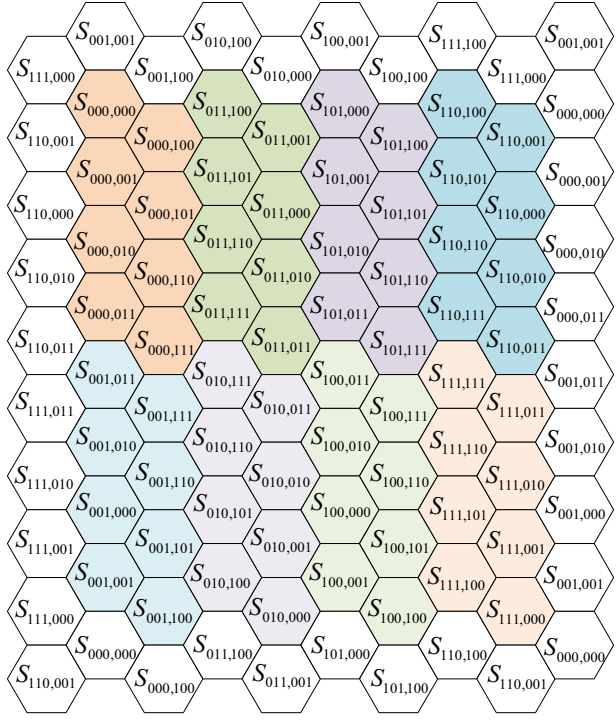}
\caption{Assignment of orthogonal sets to a lattice of regular hexagonal cells}
\label{fig:m=5}
\end{figure}
 Figure~\ref{fig:m=5} has a very regular structure depicting the use of all the 64 sequence sets $S_{c,\alpha}$ and their re-use for achieving an assignment of orthogonal sets for $D=8$. In the first place, the 8 sequence sets $S_{c,\alpha}$, for different $\alpha\in \F^3$, are placed in two adjacent columns. These sets within this cluster (marked with the same color) are clearly orthogonal using the fact that $S_{c,\alpha}\bot S_{c,\alpha'}$ when $\alpha\neq \alpha'$. A neighbouring horizontal cluster of cells $\{S_{c',\alpha} : \alpha \in \F^3 \}$ is chosen so that either $c+c'=011$ or $c+c'=110$.
 Then, to verify whether two adjacent cells $S_{c,\alpha}$ and $S_{c',\alpha'}$ are orthogonal (using (\ref{eq:orthog_check})), it is enough to check
 if $\oo{f_{011}}$ (or $\oo{f_{110}}$) and $H_{\alpha+\alpha'}$ are orthogonal using Table~\ref{tab:m=5}. For instance, considering the upper part in Figure~\ref{fig:m=5}, the sets $S_{000,\alpha}$ (in orange) and $S_{011,\alpha}$ (in green) satisfy that $c+c'=011$ and the neighbouring cells satisfy that either $\alpha=\alpha'$ (which always implies orthogonality) or $\alpha+\alpha'\in \{001,011\}$. Then, using Table~\ref{tab:m=5} again, one can verify that $\oo{{f}_{011}}$ is orthogonal to both $H_{001}$ and to $H_{011}$ which essentially implies that the neighbouring cells $S_{000,\alpha}$ and $S_{011,\alpha'}$ are orthogonal to each other.
 The orthogonality of any two adjacent cells can be readily checked using Table~\ref{tab:m=5}.
\end{example}

To fairly compare our design with the approach taken in \cite{Smith2010}, where each cell contains a set $S_{i,j,k}$ which is of cardinality $2^{m-3}$ (the number of sequences), we must use $2^{t-3}$ sets $S_{c,\alpha}$ in each cell to have the same number of sequences $2^s 2^{t-3} =2^{m-3}$. Then the number of sets decreases from $2^{2t}$ to $2^{t+3}$, and accordingly the re-use distance satisfies $D\leq \sqrt{2^{t+3}}$.
When $m>5$ ($t>3$), the problem of assigning several sets $S_{c,\alpha}$ to each cell, while preserving the orthogonality and the re-use distance of such a multiset to multisets in adjacent cells, seems to be a hard combinatorial challenge. We will avoid this hard combinatorial problem by providing alternative design methods given in the next section.

\section{Simplifying the assignment of orthogonal sets to cells}\label{sec:modif}

In order to increase the number of sequences per cell while preserving the orthogonality, two modifications of our main method are discussed in this section.

\subsection {Simplifying cell assignment through bent concatenation}

 Given large sets of orthogonal sequences of length $2^m$, through a suitable bent concatenation these sequences can be extended to any desired length $2^{m+u}$, where $u \geq 4$ is an even number. At the same time, the number of orthogonal semi-bent sequences in each $S_{c,\alpha}$ is increased by a factor $2^u$.

\begin{construction}\label{constr:2}
 Let $m$, $s$, and $t$ be three positive integers with $m=s+t$, where $s<t$. Let $F(y,x)=(f_1,\cdots,f_t)$, $F:\F^m \rightarrow \F^t$, be a vectorial plateaued function as in Construction \ref{constr:1},
 where $y\in \F^s$ and $x\in \F^t$. Let $z\in \F^u$ with $u\geq 2t$ be an even number. Let $H(z)=(h_1,\ldots,h_t)$, $H:\F^u \rightarrow \F^t$, be a vectorial bent function. We construct a vectorial plateaued
 function $G:\F^{m+u} \rightarrow \F^t$ as follows:
 \begin{align}\label{}
 G(z,y,x)&=H(z)+F(y,x)\nonumber\\
 &=(h_1(z)+f_1(y,x), \ldots, h_t(z)+f_t(y,x))
 \end{align}
 For $\alpha\in \F^t$, let
 \begin{equation}\label{eq:qc}
L'_{\alpha}=\{l'_{(\beta',\beta)}=(\beta',\beta,\alpha)\cdot (z,y,x)~|~\beta'\in \F^u, \beta\in \F^s\}.
 \end{equation}
For $c\in {\F^t}$, let
\begin{equation}
g_{c}(z,y,x)=c\cdot G(z,y,x)=h_c+f_c,
\end{equation}
where $h_c=c\cdot H$ and $f_c=c\cdot F$.
 We construct $2^{2t}$ disjoint sets of orthogonal sequences $S'_{c,\alpha}$, each of cardinality $2^{u+s}=2^{u+m-t}$, as follows:
\begin{equation}\label{eq:}
S'_{c,\alpha}=\{\oo{g_c} * \oo{l'}~|~l'\in L'_\alpha\},~~\mathrm{for}~c, \alpha\in {\F^t}.
\end{equation}
\end{construction}

\begin{theorem}
For $c, \alpha\in {\F^t}$, let the sequences sets $S_{c,\alpha}$ and $S'_{c,\alpha}$ be constructed as in Construction \ref{constr:1} and in
Construction ~\ref{constr:2}, respectively. Then
\begin{enumerate}[i)]
\item For any fixed $e, \delta\in {\F^t}$, $S'_{c,\alpha}\bot S'_{e,\delta}$ if and only if $S_{c,\alpha}\bot S_{e,\delta}$;
\item All the sequences in $S_{c',\alpha'}$ are orthogonal plateaued sequences, and $\frac{\# S'_{c,\alpha} }{\# S_{c,\alpha}}=2^u$;
\item For $t=\lfloor (m+2)/2\rfloor$, the sequences of $S'_{c,\alpha}$ are semi-bent, and the maximum correlation between two non-orthogonal sequences is equal to $\lfloor 2^{(m+u+2)/2}\rfloor$.
\end{enumerate}
\end{theorem}

\begin{proof}

$i)$ For $c, \alpha\in {\F^t}$, let
 \begin{equation}
\Gamma_{c,\alpha}=f_{c} + L_{\alpha}=\{f_{c} + l \mid l \in L_{\alpha}\}
 \end{equation}
 and
 \begin{equation}
\Gamma'_{c,\alpha}=g_{c} + L'_{\alpha}=\{g_{c} + l'\mid l' \in L'_{\alpha}\},
 \end{equation}
 where $L_{\alpha}=\{(\beta,\alpha)\cdot (y,x)~|~\beta\in \F^s\}$ and $L'_{\alpha}=\{(\beta',\beta,\alpha)\cdot (z,y,x)~|~\beta'\in \F^u, \beta\in \F^s\}.$
For $c$, $e\in {\F^t}$, $h_c+h_e=h_{c+e}$ is a bent function.
Let $g_{c,\alpha}\in \Gamma'_{c,\alpha}$ and $g_{e,\delta}\in \Gamma'_{e,\delta}$.
We have
 \begin{align}
W_{ g_{c,\alpha} + g_{e,\delta}}(\textbf{0}_{u+m})
&= \sum_{(z,y,x)\in \F^{u+m}}(-1)^{g_{c} + (\beta',\beta,\alpha)\cdot (z,y,x) + g_{e} + (\theta',\theta,\delta)\cdot (z,y,x)}\nonumber\\
&= \sum_{z\in \F^u}(-1)^{h_{c}(z)+h_{e}(z)+(\beta'+\theta')\cdot z}
\cdot \sum_{(y,x)\in \F^m}(-1)^{f_{c}(y,x) + (\beta,\alpha)\cdot (y,x) + f_{e}(y,x) + (\theta,\delta)\cdot (y,x)}\nonumber\\
&= W_{h_{c+e}}(\beta'+\theta') W_{ f_{c,\alpha} + f_{e,\delta}}(\textbf{0}_m),
 \end{align}
where $f_{c,\alpha}\in \Gamma_{c,\alpha}$ and $f_{e,\delta}\in \Gamma_{e,\delta}$.
Noticing that $ W_{h_{c+e}}(\beta'+\theta')=\pm 2^{u/2}\neq 0$, we have that
$W_{ g_{c,\alpha} + g_{e,\delta}}(\textbf{0}_{u+m}) = 0 $ if and only if $W_{ f_{c,\alpha} + f_{e,\delta}}(\textbf{0}_m) = 0$.
Note that $ S_{c,\alpha}=\{\oo{{(f_c + l)}}~|~l\in L_\alpha\} $ and $S'_{c,\alpha}=\{\oo{{(g_c + l')}}~|~l'\in L'_\alpha\}$.
By Lemma \ref{orth}, $S'_{c,\alpha}\bot S'_{e,\delta}$ if and only if $S_{c,\alpha}\bot S_{e,\delta}$.

$ii)$ Let $\beta'\in \F^u$, $\beta\in \F^s$, and $\alpha\in \F^t$. We have
\begin{align}
 W_{g_c}(\beta',\beta, \alpha) &= \sum_{(z,y,x)\in \F^{u+m}}(-1)^{g_c+(\beta',\beta, \alpha)\cdot (z,y,x)}\nonumber \\
 &= \sum_{z\in \F^u} (-1)^{h_c+\beta'\cdot z}\sum_{(y,x)\in \F^m}(-1)^{f_c+(y.z)\cdot (\beta, \alpha)} \nonumber \\
 &= W_{h_c}(\beta')W_{f_c}(\beta,\alpha)
\end{align}
Note that $h_c$ is a bent function and $f_c$ is a plateaued function. More precisely, $W_{h_c}(\beta')\in \{\pm 2^{u/2}\}$ and $W_{f_c}(\beta,\alpha)\in \{0, \pm 2^t\}$.
Then we have $ W_{g_c}(\beta',\beta, \alpha)\in \{0, \pm 2^{u/2 + t}\}$, which implies $g_c$ is a plateaued function. Furthermore, $g_c + l'$ is also a plateaued function,
where $l'\in \B_{u+m}$ is a linear function. Then all the sequences in $S'_{c,\alpha}=\{\oo{{(g_c + l')}}~|~l'\in L'_\alpha\}$ are plateaued sequences. Let $\oo{{(g_c + l_1')}}$ and $\oo{{(g_c + l_2')}}$ be any two different sequences in $S'_{c,\alpha}$.
Noticing $(g_c + l_1')+(g_c + l_2')=l_1'+l_2'$ is a balanced function, $S'_{c,\alpha}$ is a set of orthogonal sequences. In addition,
$\frac{\# S'_{c,\alpha} }{\# S_{c,\alpha}}=\frac{\# \Gamma'_{c,\alpha} }{\# \Gamma_{c,\alpha}}=\frac{\# L'_{c,\alpha} }{\# L_{c,\alpha}}=2^u$.

$iii)$ In particular, when $t=\lfloor (m+2)/2\rfloor$, the sequences $g_c + l'$ (for $l'\in L'_\alpha$) are semi-bent, which implies that
the sequences of $S'_{c,\alpha}$ are semi-bent, and the maximum correlation between any two non-orthogonal sequences is equal to $2^{u/2}\cdot 2^t=\lfloor 2^{(m+u+2)/2}\rfloor$, as claimed.
\end{proof}

\begin{remark}
Notice that the increase of cardinality of the sets $S_{c,\alpha}$ by a factor $2^u$ cannot be achieved by Construction~\ref{constr:1}. Indeed, considering semi-bent sequences, we may take $m'=m+u=s'+t'$ in Construction~\ref{constr:1}, where for odd $m$ and even $u$ we necessarily have $t'=s'+1$. For instance, taking $m=5$ and $u=6$ Construction~\ref{constr:2} gives $2^{s+u}=2^8$ orthogonal sequences within each $S_{c',\alpha'}$, whereas using Construction~\ref{constr:1} with $m'=11$ so that $s'=5$ and $t'=6$ only $2^{s'}=2^5$ orthogonal sequences within each $S_{c,\alpha}$ are obtained. Thus, Construction~\ref{constr:1} implies a greater combinatorial challenge than Construction~\ref{constr:2} since we need to assign $2^3$ orthogonal sets $S_{c,\alpha}$ within each cell in order to attain the same number of users (sequences).
\end{remark}

\subsection{Assignment of orthogonal sets for $D=4$}

For practical applications, there are many indications that the re-use distance $D=4$ is quite sufficient.
Indeed, inspection of
an antenna polar diagram for a satellite system (describing the imperfect
directionality of an antenna) suggests that $D=4$ is an adequate re-use distance for both a regular hexagonal tessellation and for an irregular network (cf. \cite{Smith2013} for assignment of codewords to irregular networks). A similar comment arises for a quasi-synchronous
CDMA terrestrial system by considering the propagation loss. Therefore, our main goal is to solve the problem of assigning $2^{m-2}$ sequences per cell when $D=4$, regardless of the parity and size of $m$.

\subsubsection{$m=3$ and other odd cases}\label{sec:m=3}
To cover the case $m$ odd, it is suitable to take the smallest odd $m=3$ in Construction~\ref{constr:1} for which an arrangement of cells with $D=4$ is possible. The whole procedure of specifying the sets $S_{c,\alpha}$ and their exact placement within a regular hexagonal structure is given in the example below.
\begin{example}\label{ex:m=3}

Let $m=3$, $s=1$, and $t=2$. Let $\gamma\in \FF_{2^2}$ be a root of the primitive polynomial $z^2+z+1$.
We have $\pi(1)=10$, $\pi(\gamma)=01$, $\pi(\gamma^2)=11$, $\pi(0)=00$.
For $y\in \F$, $x=(x_1,x_2)\in \F^2$, a vectorial semi-bent function $F: \F^{3} \mapsto \F^{2}$ is constructed as
$F(y,x)=(f_1, f_2),$
where
\begin{eqnarray*}
 f_1(y,x) &=& \pi(\gamma^{[y]+1})\cdot x
=(y+1)x_1 +y(x_1+x_2),
\\
 f_2(y,x) &=& \pi(\gamma^{[y]+2})\cdot x
=(y+1)(x_1+x_2) +yx_2.
\end{eqnarray*}
Let $f_c(y,x)=c\cdot F(y,x)$, for $c\in \F^3$. For any $\alpha \in \F^2$, a subset of Hadamard sequences can be defined as
 $H_{\alpha}=S_{00,\alpha}=\{\oo{l}~|~l\in L_\alpha\}.$
 The sets of sequences sets $\{S_{00,\alpha}~|~\alpha\in \F^2\}$ then correspond to a partition of the Hadamard matrix $\mathcal {H}_3$ into 4 parts.
 In Table~\ref{tab:m=3}, we depict the orthogonality between $\overline{f_{c}}$ and $H_{\alpha}$.
Then, by Theorem~\ref{th1}, $S_{c,\alpha}\bot S_{c',\alpha'}$ if and only if one of the following conditions holds:
\begin{itemize}
 \item $c=c'$ and $\alpha\neq \alpha'$
 \item $c\neq c'$ and $\alpha=\alpha'$
 \item $c+c'=\alpha+\alpha'$ when $c\neq c'$ and $\alpha\neq \alpha'$.
\end{itemize}
The re-use distance is $D=4$ according to the arrangement given in Figure~\ref{fig:m=3}.
\begin{table}[t]
{
\caption{Orthogonality between $\oo{{f_{c}}}$ and $H_{\alpha}$ in Example \ref{ex:m=3} }
\label{tab:m=3}
\begin{center}
\begin{tabular}{|c| c| c| c| c| c| c| c| c|c|c| }
\hline
 &$H_{00}$ &$H_{10}$ &$H_{01}$ &$H_{11}$\\ \hline
$\oo{f_{00}}$& & $\bot$ & $\bot$ & $\bot$ \\\hline
$\oo{f_{10}}$& $\bot$ & $\bot$ & & \\\hline
$\oo{f_{01}}$& $\bot$ & & $\bot$ & \\\hline
$\oo{f_{11}}$& $\bot$ & & & $\bot$ \\\hline
\end{tabular}
\end{center}}
\end{table}

\begin{figure}[t]
\centering
\includegraphics[scale=0.6]{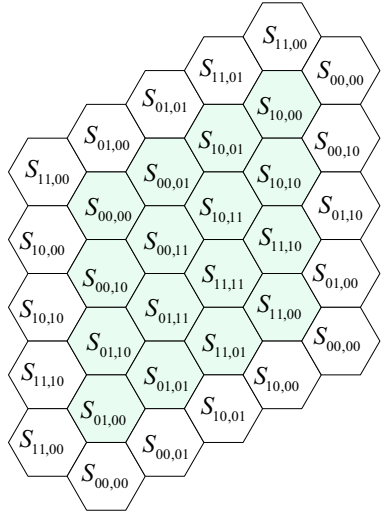}
\caption{Assignment of orthogonal sets to a lattice of regular hexagonal cells, $D=4$}
\label{fig:m=3}
\end{figure}
\end{example}

By employing Construction \ref{constr:2} with $m=3$, through the parameter $u$ the assignment of $2^{m'-2}$ sequences per cell can be achieved for $D=4$ and for any odd $m' \geq 7$, where $m'=m+u$.

\subsubsection{$m=6$ and other even cases}\label{sec:single}

In order to apply a similar approach for resolving the problem of assigning orthogonal sets when $m$ is even, we specify a modified design method in this subsection.
 The main reason for this is the non-efficiency (in terms of the number of users per cell) when using the same technique applied to the $m$ odd case. Indeed, using Construction~\ref{constr:1} for even $m=s+t$ (where $s=m/2-1$ and $t=m/2+1$ is necessary to obtain semi-bent sequences) we would in general have $2^{2t}=2^{m+2}$ many sets $S_{c,\alpha}$ and each set would contain $2^s=2^{m/2-1}$ sequences. Then, for any even $m\geq 4$, a large cardinality of $\{S_{c,\alpha}\mid c,\alpha \in \F^t\}$ would give a relatively low number of sequences in each $S_{c,\alpha}$, i.e., $\# S_{c,\alpha}=2^{m-t}$, where $t\geq 3$.

\begin{construction}\label{constr:3}
Let $m$, and $k$ be two positive integers with $m=2k+2$. Let $\gamma$ be a primitive element of $\FF_{2^k}$, and $\{1,\gamma, \ldots, \gamma^{k-1}\}$ be a polynomial basis of $\FF_{2^k}$ over $\F$.
Define the isomorphism $\pi$: $\FF_{2^k}\mapsto \F^k$ by
 \begin{equation*}
 \pi(b_1+b_2\gamma+\cdots+b_{k}\gamma^{k-1})=(b_1,b_2,\ldots,b_{k}).
 \end{equation*}
For $i=1, \ldots,k$, let $\phi_i:\F^k \rightarrow \F^k$ be a bijective mapping defined by
\begin{eqnarray}
\phi_i(y)=\left\{\begin{array}{ll}
 \textbf{0}_{k}, & y=\textbf{0}_{k}\\
 \pi(\gamma^{[y]+i}), & y\in {\F^k}^*
\end{array} \right.
\end{eqnarray}
where $[y]$ denotes the integer representation of $y$.
Let $y\in \F^k$, $x\in \F^{k+2}$. For $i=1, \ldots,k$, let
 \begin{equation}\label{32}
f_i(y,x)=(\phi_i(y),00)\cdot x.
 \end{equation}
We define a semi-bent vectorial function $F: \F^{m} \mapsto \F^{k}$ by
 \begin{equation}
F(y,x)=(f_1,\ldots, f_k).
 \end{equation}
Let $d\in \{2,3\}$ and for any fixed $\alpha\in \F^d$ define
 \begin{equation}
L_{\alpha}=\{l_\beta=(\beta,\alpha)\cdot (y,x)~|~ \beta\in \F^{m-d}\}.
 \end{equation}
For any $c\in {\F^k}$, let $f_{c}(y,x)=c\cdot F(y,x)$. We construct $2^{k+d}$ disjoint sets of sequences each of cardinality $2^{m-d}$ as follows:
\begin{equation}\label{eq:Sc3}
S_{c,\alpha}=\{\oo{f_c} * \oo{l}~|~l\in L_\alpha\},~~\mathrm{for}~c\in {\F^k},~\alpha\in \F^d.
\end{equation}
\end{construction}

\begin{theorem}\label{th3}
For $c\in {\F^k}$, $\alpha\in \F^d$, let the sets of sequences $S_{c,\alpha}$ be defined by (\ref{eq:Sc3}) as in Construction \ref{constr:3}.
Then, we always have
\begin{enumerate}[i)]
\item All the sequences in $S_{c,\alpha}$ are semi-bent sequences;
\item For any $c\in {\F^k}$, $\alpha\in \F^d$, $S_{c,\alpha}$ is a set of orthogonal sequences with $\#S_{c,\alpha}=2^{m-d}$;
\item Let $c\in {\F^k}$, $\alpha=(\alpha_1,\ldots,\alpha_d)\in \F^d$.
 For any fixed $e \neq c \in {\F^k}$, $\delta=(\delta_1,\ldots,\delta_d)\in \F^d$,
 $S_{c,\alpha}\bot S_{e,\delta}$ if and only if $(\alpha_{d-1},\alpha_d) \neq (\delta_{d-1},\delta_d)$.
 In addition, $S_{e,\alpha}\bot S_{e,\delta}$ if and only if $\alpha\neq \delta$;
\item There exist $3\cdot 2^{k+d-2}+2^{d-2}-1$ many sets in
 $\Omega=\{S_{c,\alpha}~|~c\in {\F^k},\alpha\in \F^d\}$ that are orthogonal to $S_{e,\delta}$.
\end{enumerate}
\end{theorem}

\begin{remark}\label{rm5}
The above results can be easily deduced using exactly the same techniques as in the proof of Theorem~\ref{th1}. More precisely, one can easily show that when $c\neq e$ then there exist $3\cdot 2^{k+d-2}$ many sets in $\Omega$ that are orthogonal to $S_{e,\delta}$. On the other hand, when $c = e$ there are $2^{d-2}-1$ many sets in $\Omega$ orthogonal to $S_{e,\delta}$.
\end{remark}

Since each $S_{c,\alpha}$ is of cardinality $2^{m-d}$, taking $d=2$ implies that each cell is assigned a single set $S_{c,\alpha}$ with $2^{m-2}$ orthogonal sequences (twice as many as the method in \cite{Smith2010}), thus making the arrangement of the remaining orthogonal sets in (non)adjacent cells much easier.
Notice that taking $d=3$ would give us the possibility of assigning $2^{m-3}$ sequences per cell with larger re-use distance.

\begin{example}\label{ex:m=6}
Let $m=6$. Let $\gamma\in \F^2$ be a root of the primitive polynomial $z^2+z+1$.
Then, using $\pi(\textbf{0})=00$, $\pi(1)=10$, $\pi(\gamma)=01$, $\pi(\gamma^2)=11$,
let
$\phi_1(00)=\pi(0)$, $\phi_1(01)=\pi(\gamma^2)$, $\phi_1(10)=\pi(1)$, $\phi_1(11)=\pi(\gamma)$;
$\phi_2(00)=\pi(0)$, $\phi_2(01)=\pi(1)$, $\phi_2(10)=\pi(\gamma)$, $\phi_2(11)=\pi(\gamma^2).$
For $y=(y_1,y_2)\in \F^2$, $x=(x_1,x_2,x_3,x_4)\in \F^4$, a vectorial semi-bent function $F: \F^{6} \mapsto \F^{2}$ is constructed as
 $F(y,x)=(f_1, f_2),$
where
\begin{eqnarray*}
 f_1(y,x) &=& (\phi_1(y),00)\cdot x\\
&=&\oo{y}_1\oo{y}_2\cdot 0
+\oo{y}_1y_2(x_1+x_2)
+{y}_1\oo{y}_2 x_1
+y_1y_2 x_2,
\\
 f_2(y,x) &=& (\phi_2(y),00)\cdot x\\
&=&\oo{y}_1\oo{y}_2\cdot 0
+\oo{y}_1y_2 x_1
+{y}_1\oo{y}_2 x_2
+y_1y_2(x_1+x_2). \\
\end{eqnarray*}
Let $f_c(y,x)=c\cdot F(y,x)$, for $c\in \F^2$.
Taking $d=2$, we have $\# \{S_{c,\alpha}~|~c\in \F^k, \alpha\in \F^d\}=2^{k+d}=16$ and each $S_{c,\alpha}$ is of cardinality $2^{m-d}=16$. For any $\alpha \in \F^2$, let $H_{\alpha}=\{\oo{{{l_{\beta}}}} \mid l_{\beta} \in L_{\alpha} \} = S_{00,\alpha}$. Note that $\{H_{\alpha} :
\alpha \in \F^2\}$ corresponds to a partition of the Hadamard matrix $\mathcal {H}_6$ into 4 parts. In Table~\ref{tab:m=6}, we depict the orthogonality between $\oo{{f_{c}}}$ and $H_{\alpha}$.
\begin{table}[t]
{
\caption{Orthogonality between $\oo{f_{c}}$ and $H_{\alpha}$}
\label{tab:m=6}
\begin{center}
\begin{tabular}{|c| c| c| c| c| c| c| c| c|c|c| }
\hline
 &$H_{00}$ &$H_{01}$ & $H_{10}$& $H_{11}$ \\ \hline
$\oo{{f}_{00}}$& & $\bot$ & $\bot$ & $\bot$ \\\hline
$\oo{{f}_{01}}$& & $\bot$ & $\bot$ & $\bot$ \\\hline
$\oo{{f}_{10}}$& & $\bot$ & $\bot$ & $\bot$ \\\hline
$\oo{{f}_{11}}$& & $\bot$ & $\bot$ & $\bot$ \\\hline

\end{tabular}
\end{center}}
\end{table}
Note that for $c\in {\F^k}$, $\alpha\in \F^d$, we have
$S_{c,\alpha}=\{\oo{f_c}*\oo{l}~|~\oo{l}\in H_\alpha\}
 =\{\oo{f_c + l}~|~l\in L_\alpha\}.$
\begin{figure}[t]
\centering
\includegraphics[scale=0.6]{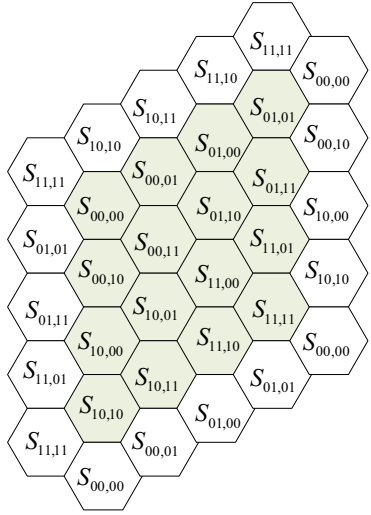}
\caption{Assignment of orthogonal sets to a lattice of regular hexagonal cells}
\label{fig:m=6}
\end{figure}
Thus, $S_{c,\alpha}\bot S_{c',\alpha'}$ if and only if $\alpha \neq \alpha'$.
It is very easy to give an arrangement of the 16 sets $S_{c,\alpha}$ with the re-use distance $D=4$, as shown in Figure \ref{fig:m=6}. Obviously, the sets in adjacent cells are orthogonal.
\end{example}

Now using Construction~\ref{constr:3} with $m=6$ and the fixed arrangement of the sets $S_{c,\alpha}$ given in Figure~\ref{fig:m=6}, the problem of assigning $2^{m-2}$ sequences to each cell is resolved for any even $m\geq 10$ (including $m=6$) by applying Construction \ref{constr:2}.

\subsection{Comparison to other designs} \label{sec:comp}

In Table~\ref{tab:c2}, we compare our design methods to the approach taken in \cite{Smith2010} in terms of the number of users per cell $\mathcal{N}$. In addition, we also list the cardinalities of the sets of orthogonal semi-bent sequences $\#\Omega=\# \{S_{c,\alpha}~|~c,\alpha\in \F^t\}$. While the assignment of $\mathcal{N}=2^{m-2}$ users per cell was only possible for $m=3,9$ in \cite{Smith2010, Smith2012} when $D=4$ (otherwise the number is $\mathcal{N}=2^{m-3}$), using our methods this assignment is achievable for any $m \neq 4,5$.

\begin{table}[t]
{
\caption{A comparison of the main parameters}
\label{tab:c2}
\begin{center}
\begin{tabular}{|c| c| c| c| c| c| c| c| c|c|c| }
\hline
Methods &$\N$ &$\#\Omega$ &$D$ \\ \hline
\footnotesize{One Hadamard Matrix} &$2^{m-2}$ &4 & 2 \\\hline
$\textrm{Smith {\em et al.} \cite{Smith2010,Smith2012}}\atop {(m=3,9)}$ &$2^{m-2}$ &16 & 4 \\\hline
$\textrm{Cons. \ref{constr:1} } \& \textrm{ Cons. \ref{constr:2} } \atop {\textrm{($m\geq 3$ odd and $m\neq 5$)}}$ &$2^{m-2}$ &16 & 4 \\\hline
$\textrm{Cons. \ref{constr:3} } \& \textrm{ Cons. \ref{constr:2} } \atop {(\textrm{$m\geq 6$ even})}$ &$2^{m-2}$ &$2^{m/2+1}$&$\geq 4$\\\hline
\end{tabular}
\end{center} }
\end{table}

\section{Quasi-orthogonal sequences and the window property}\label{sec:qos}

Quasi-orthogonal sequences (QOS) were introduced in \cite{Yang2000} as a means of increasing the number of channels in synchronous CDMA systems. The orthogonality is traded-off against the increased capacity of the system, though there are some other desirable features such as the {\em window property} which characterize this family of sequences. The notion of the covering radius of the first order Reed-Muller code is quite useful here. Denoted by $\Theta_{\textnormal{min}}(N)$, where $N=2^m$ is the length of codewords, it specifies the minimum achievable correlation of any function (sequence) in $\B_m$ to the set of linear functions $\mathcal{L}_{m}$. More precisely, $\Theta_{\textnormal{min}}(N)= \min_{f \in \B_m} \max_{\ell \in \mathcal{L}_m} |W_{f+\ell}(\textnormal{{\bf 0}}_m)| $. It is well-known that for even $m$ the minimum value $\Theta_{\textnormal{min}}(N)=2^{m/2}$ is achieved by bent functions, whereas for odd $m$ the exact value of $\Theta_{\textnormal{min}}(N)$ is not determined for $m\geq 9$ but in general it satisfies $2^{m/2} < \Theta_{\textnormal{min}}(N) \leq 2^{(m+1)/2}$.

A family $\mathcal{G}=\{g_i(x): i=1, \ldots, M \}$
of functions (sequences) of length $N=2^m$ is said to be quasi-orthogonal if $\mathcal{G}$ contains $\mathcal{L}_m$, $|W_{g_i+g_j}(\textnormal{{\bf 0}}_m| \leq \Theta_{\textnormal{min}}(N)$ for any two sequences $g_i,g_j \in \mathcal{G}$, and any subsequence of length $2^v$ (dividing any sequence of length $2^m$ in $\mathcal{G} \setminus \mathcal{L}_m$ into $2^{m-v}$ subsequences) achieves the minimum possible correlation related to $\mathcal{L}_v$, where $2 \leq v \leq m$.

The first two conditions refer to the minimum achievable correlation for original length, whereas the last condition is known as the window property and is motivated by practical applications for achieving flexible data rates without violating the minimum correlation property. For instance, instead of transmitting the codewords of length $2^m$ the transmission rate can be doubled by sending codewords of length $2^{m-1}$ and at the same time good (minimum) correlation properties of the shortened sequences are preserved. Notice that for odd $m$, the value of $\Theta_{\textnormal{min}}(N)$ is taken to be $2^{(m+1)/2}$ which essentially corresponds to semi-bent sequences.

Another similar concept, known as orthogonal variable spreading factor (OVSF) codes \cite{Dinan98,Smith2012} that are commonly employed in W-CDMA (wideband CDMA) systems, greatly coincides with the notion of QOS. Roughly speaking, the main conditions of keeping the correlation as low as possible even though the original sequences of length $2^m$ are divided into $2^{m-v}$ blocks of length $2^v$ are valid for OVSF codes as well. The main difference to QOS is that OVSF codes require the orthogonality of sequences in adjacent cells and therefore sequences stemming from bent functions along with the Hadamard sequences cannot be used in this case. Note that this is a condition required in \cite{Smith2012} and is also used in this work, but it is not part of the standard definition of OVSF codes.

In what follows, we show that Construction~\ref{constr:2} may potentially give rise to OVSF codes satisfying the window property but only partially.
Without going into technical details we discuss the properties of Construction~\ref{constr:1} with respect to OVSF codes. In the first place, the first two properties are satisfied since obviously $\mathcal{L}_{m} \in \{S_{c,\alpha}\}$ (namely $\mathcal{L}_{m} = S_{{\bf{0}},\alpha}$) and the sequences generated by Construction~\ref{constr:1} are semi-bent sequences. However, the window property is generally not satisfied which is easily confirmed by setting the variable $y \in \F^s$ to any fixed value. Indeed, the function $f_i(y,x)=\phi_i(y)\cdot x$ in Construction~\ref{constr:1} is just a linear function and the minimum correlation cannot be achieved.

The window property (along with the orthogonality in adjacent cells) being hard to satisfy in general, the authors in \cite{Smith2012} introduced the concept of {\em semi-bent depth} of order $r$. A function $g \in \B_m$ is said to satisfy semi-bent depth of order $r$ if all its restrictions of length $2^{m-s}$, where $0 \leq s \leq r$, are semi-bent functions in
$m - s$ variables. Suitable semi-bent functions, having semi-bent depth approximately equal to $r=(m-1)/2$, were found in \cite{Smith2012} using a computer search.
Our Construction~\ref{constr:2} gives semi-bent sequences of length $2^{m+u}$ by extending semi-bent sequences of length $2^m$ through bent concatenation. Thus, all the conditions above are satisfied for the sequences of length $2^m$ and $2^{m+u}$.
However, there is no guarantee that the sequences of length $2^{m+u-r}$, for $1 \leq r \leq u-1$, are semi-bent.

To fully satisfy the window property of semi-bent depth of order $u$, the vectorial bent function $H$ used in Construction~\ref{constr:2} should be such that its linear combinations of the component functions satisfy the property that their subfunctions (subsequences) are always semi-bent. This question is however left as an interesting open problem.

\section{Conclusions} \label{sec:conc}

In this paper, concerning the assignment of orthogonal sequences to a regular tessellation of hexagonal cells, we have shown that in most of the cases a larger number of users per cell (twice as many) can be assigned than using previously known methods. In particular, for $D=4$ and any $m \geq 3$ (with $m \neq 4,5$) the number of users (sequences) per cell is equal to $2^{m-2}$, for the codewords of length $2^m$.

%

\end{CJK*}

\end{document}